Investigating Deletion in Wikipedia

Bluma S. Gelley

Polytechnic Institute of NYU

January 11, 2013

## 1.1 Abstract

Several hundred Wikipedia articles are deleted every day because they lack sufficient notability to be included in the encyclopedia. We collected a dataset of deleted articles and analyzed them to determine whether or not the deletions were justified. We find evidence to support the hypothesis that many deletions are carried out correctly, but also find that a large number were done very quickly. Based on our conclusions, we make some recommendations to reduce the number of non-notable pages and simultaneously improve retention of new editors.

## 1.2 Introduction

Wikipedia is perhaps one of the largest examples of open collaboration on the Web today. Its size, diversity, and the richness of the data it records make it an ideal object of study for those wishing to investigate any aspect of collaborative work. One important aspect is the processes surrounding the removal of irrelevant content, and the relation this has to the retention of new collaborators.

## 1.3 Overview

Wikipedia, the 6[th] most visited web site in the world [1], is entirely community-created. The English-language version of the encyclopedia has 4.1 million articles and over 18 million registered users. Anyone, registered or not, can edit any article by simply clicking on the "Edit" link at the top of the page and entering text. (Throughout this article, we use the terms "user" and "editor" interchangeably.)Users can also remove, or revert, each other's edits with a single click. To create an article, an account is required; registration, though, is very simple, and pages are visible on the site as soon as they are created. Wikipedia has developed into a complex ecosystem, with community-created rules, called policies, and an intricate social system. Despite frequently voiced doubts about its reliability, it is widely used as an information resource and has been proven to be nearly as accurate as Encyclopedia Britannica [1].

Like any other encyclopedia, Wikipedia has rules for what topics it deems worthy of inclusion. The site is very clear as to what it is not: Wikipedia is not a blog, a social network, or a directory, and it is definitely not "an indiscriminate collection of information" [2]. It is an encyclopedia, albeit one with more available space than a traditional one. Though it welcomes articles on topics not generally included in a print encyclopedia, it still maintains standards of notability to ensure that all topics covered are fundamentally *encyclopedic*. Namely, the "topic has received significant coverage in reliable sources that are independent of the subject" and that are easily verifiable [3]. Additional specific notability guidelines are voluminous and extremely detailed, with policies governing the notability of academics, books, and organizations, among others. The ease with which articles can be created necessitates a method for removing the many unencyclopedic articles that are bound to result. Among Wikipedia's policies, therefore, are many governing the process of article deletion, a complex system that attempts to streamline the deletion process as much as possible while maintaining the site's collaborative ethos.

**2.1 Article Deletion**

Like all decisions on Wikipedia, article deletion is a collective one. Only administrators can delete articles, but any user can nominate an article for deletion. There are three tiers of deletion procedures. The lowest, Speedy Deletion, is for pages which should so obviously be removed that there is no need for discussion. There are several dozen criteria for speedy deletion, including technical deletions (e.g. as part of moving a page), vandalism, blatant advertising, and articles that make no claim to notability. Speedy deletion candidates can be, and often are, deleted within minutes of their nomination.

The next tier is Proposed Deletion, or PROD, for articles that clearly don't belong, but don't fit the Speedy criteria. PRODs are subject to a 7-day waiting period; if anyone contests the deletion during that time, the article is de-PROD'ed. It can, however, be re-nominated for Speedy Deletion, or pushed up a level to AFD. There is also a separate PROD process for new biographies of living people containing no sources.

AfD, or deletion discussion, is the highest tier. Any article which is not as blatantly unencyclopedic as a Speedy or a PROD is brought here for community discussion. Users leave comments with their recommendations for whether it should be kept or deleted, and an administrator carries out the action chosen by general consensus.

Pages that are deleted are removed from the site. The only public record is kept in the deletion log, containing basic metadata about the deletion itself and its reason, but none about the actual page.

(For a flowchart illustrating the deletion process, see [4].)

**2.2 Some statistics**

To understand the importance of deletion to the smooth operation of Wikipedia, it helps to look at some numbers. Between 1,000 and 1,500 articles are created on the English Wikipedia every day. Of these, some 20% are deleted within a week, most of these by Speedy Deletion. In total, about 1000 articles are deleted every day. 37% percent of all speedy deletions are for lack of notability [5].

Once an article is nominated for deletion, it is quite likely to be deleted. About 75% of PRODs and 80% of Speedy nominees are deleted. The huge numbers of daily deletions and their strategic importance to Wikipedia's operation makes deletion an interesting topic of study.

**3.1 Related Work**

Most previous research on deletion in Wikipedia has focused on analyzing various aspects of deletion discussions [4] [6] [7]. No one has taken more than a superficial look at Speedy Deletion, the most common type of deletion and, due to the lack of community discussion, the one with the greatest potential for abuse. One exception is [8], who attempted to determine if deleted article topics were actually less notable than kept ones. However, they had to use very rough metrics since they had no access to the deleted pages themselves and were forced to rely on the deletion log; the results were inconclusive. [8] measured the percentage of speedy deletions where the rationale given was lack of notability, but they, too, relied only on the deletion log and were unable to determine whether the deleted articles were actually less notable. We decided to attempt an in-depth review of Speedy Deletion to determine whether the deletion guidelines, those about notability in particular, are being applied correctly.

**3.2 Motivation**

We were motivated by two problems troubling Wikipedia. The first is the large number of inappropriate articles created on Wikipedia. Tracking down and removing these articles eats up enormous amounts of manpower and time that could be spent improving the quality of Wikipedia. Not only does inappropriate content lower the quality, and therefore the credibility, of Wikipedia, it also impedes improvements to that quality, since users who would otherwise be working on articles are instead busy removing the damaging content. This content includes advertising, hoaxes, and other forms of vandalism, but the largest category of problematic pages, comprising a full 45% of non-technical deletions [5], are articles which appear to have been written in good faith, but fail to meet notability standards.

This serious drain of Wikipedia's resources is exacerbated by another problem – new editors are not being retained. While there is a steady influx of new registrations, very few go on to become active editors [9].The Wikipedia community is extremely concerned about the number of new editors who leave, and there has been extensive research on the reasons for their abandonment of the site [8] [10] [11]. Ironically, deletion may be contributing to the problem. The aggressiveness with which their good-faith page creations are deleted is one major reason cited by new editors for leaving the site [13]. Experienced Wikipedians, weary of fighting vandals, tend to have a "shoot first, ask questions later" attitude towards newly created pages, and vast numbers are deleted within minutes of creation, often before much content is added. New editors whose changes are reverted by experienced ones are much less likely to continue editing [14]; among users whose first edit is creating an article, those whose articles are deleted are more than 6 times more likely to leave immediately than those whose articles are kept [10].

Our study quickly made clear just how common non-notable deletions were. We found that 20 – 25% of all new pages created were subsequently deleted, many within a few hours or even minutes of their addition to the site, a large percentage for notability reasons. For a site that is so clear about its notability standards, why do there seem to be so many non-notable articles? Is this the fault of the page creators for flaunting policy and creating articles that are unencyclopedic? Or are the Wikipedians who delete pages too trigger-happy, incorrectly removing articles that actually do deserve to be included? Are the creators of non-notable articles promising future editors who should be reached out to, or incorrigible nuisances barely better than vandals?

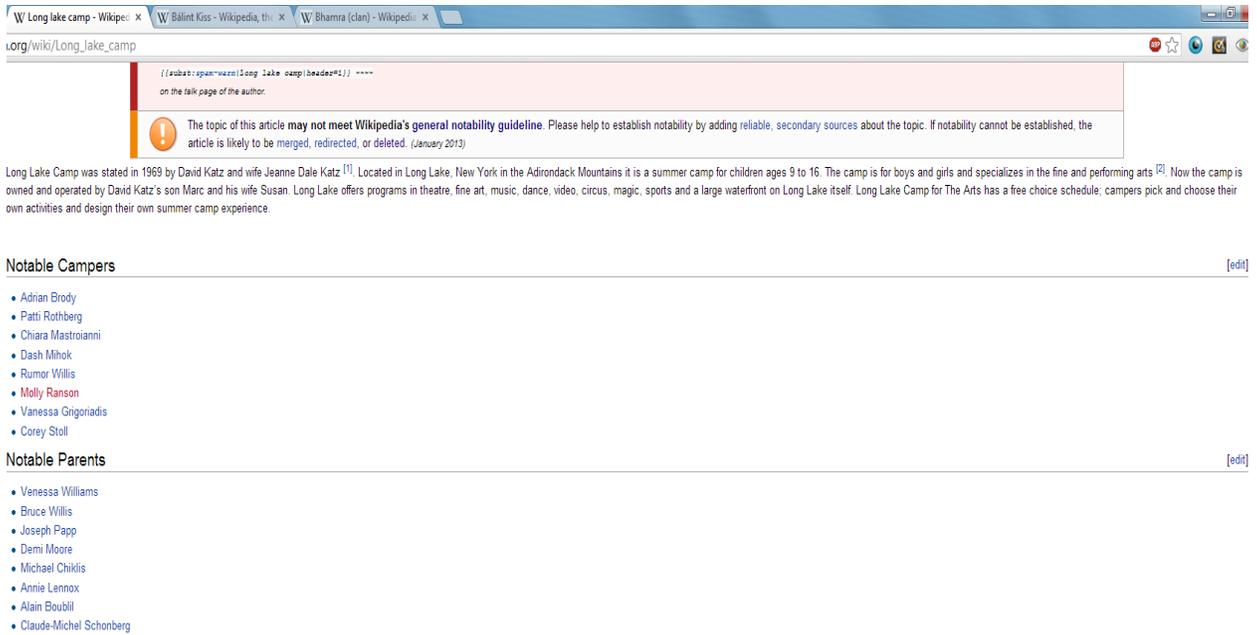

**Figure 1: An article nominated for speedy deletion. Should this topic be included in Wikipedia?**

### 3.2.1 A Closer Look

Wikipedia tries hard to address each of these problems. Great efforts are made to ensure that only encyclopedic articles are added to the site. In 2005, Wikipedia banned anonymous editors from creating pages, in an attempt to make spamming slightly less effortless. An Article Wizard walks editors through the process of creating an article, emphasizing repeatedly the need for notability and reliable sources, and then sends the article for review by experienced editors to ensure it meets Wikipedia standards. Use of the Wizard is optional, but attempting to create a page without it prompts a strong suggestion that the user read a how-to essay which is mostly comprised of warnings about what does and does not belong on the site [15]; the essay receives about 1,000 pageviews a day. The New Page Patrol checks every page created to ensure that it meets Wikipedia's standards. Experienced Wikipedians are always available to help any user with questions about the process.

To ensure that deletions are all legitimate, Wikipedia provides huge volumes of information to editors involved in deletion, which cover every possible aspect of the process, explaining in minute detail exactly what constitutes notability. New Page Patrollers are enjoined to give seemingly non-notable pages time to develop before flagging them for deletion. The criteria for allowing an article to be unilaterally deleted without consensus are extremely well-delineated and limited in scope; the slightest doubt is enough to require a week-long community discussion.

The existence of safeguards, however, does not mean that those safeguards are effective. There is often a wide gap between stated procedure and its implementation. To answer our questions from section 3.2, we must take a closer look at how deletion is actually being carried out. Manual review of each deletion nomination is unfeasible, given the huge number of deletions and the complexity of the decision regarding each one. Instead, we examine some salient characteristics of each article. If the deleted pages share common features, and these differ significantly from those of pages that are not deleted, we can infer that the article deleters are indeed doing their job correctly and the problem lies with the creators. Conversely, if there is no significant difference between the regular and deleted articles, it may signify overzealousness in the application of the notability guidelines to deletion. However, the fact that pages that are deleted from Wikipedia are no longer publicly accessible means that researchers attempting to answer this question have focused on examining the deletion log, which records only basic metadata regarding the deletion, and nothing about the page itself [8] [5]. To remedy this situation, we assembled a large dataset of deleted pages. We extracted several dozen features from each one and used them to compare the deleted pages to a set of pages that had never been deleted. The results of that comparison are detailed below.

## 4. Methods

### 4.1 Data

Collecting the data presented a challenge. It is impossible to know which pages will be deleted until they actually are deleted. On the other hand, once a page is deleted, it is no longer publicly available, and once a page is flagged as a candidate for Speedy Deletion, it can be removed at any time. To circumvent this issue, we visited the Candidates for Speedy Deletion page approximately every 12 minutes and retrieved any new entries. We then used Wikipedia's page export function to download these pages. Several days later, we checked each page to determine whether or not it had been deleted.

Using this method from October through December 2011 yielded approximately 7,300 Speedy Deletion candidate articles. A significant portion of these are technical deletions, and were excluded. We focused primarily on pages whose deletion rationale was lack of notability, but we did include articles deleted for being advertising or certain forms of spam, since these criteria are difficult to distinguish from notability problems and are frequently used interchangeably. After choosing only the pages that fit these criteria and filtering out pages that were nominated for deletion but not deleted, the final dataset contained 2,495 Speedy Deletions. Each page in the dataset contains its full edit history, including every revision ever made to the text and various metadata about each revision.

For comparison purposes, we collected a set of articles that were not deleted, and therefore presumed to be notable. We chose articles that were as similar in topic as possible to the deleted pages, so any differences between the two sets would result from notability differences, rather than topic. We purposely chose many stubs (short, underdeveloped articles) to further increase the similarity to the deleted pages, which are mostly short and incomplete. Once we had selected the topics, we used Wikipedia's category hierarchy to download the full revision history of every page in the topic. This resulted in 1,386 articles that were not deleted – the kept set.

Pages tend to be deleted soon after they are created [8]. The deleted pages in our set, therefore, are mostly very young. Since many of our features are affected by the age of the page, we attempted to control for the effect of time by creating 3 additional datasets.

-All pages, deleted or kept, older than one week,

- A "time capsule" dataset of the kept pages as they looked at 40 minutes old, along with all the deleted pages.

- A set of ~14000 new pages downloaded over 4 weeks in December 2012. Some of them were later deleted and the rest kept; all are of approximately equal (very young) age.

By equalizing the age of the pages in these three sets, we effectively normalized by time and prevented the age of the pages from becoming a confounding factor. A full description of all datasets and collection methods can be found on our webpage, http://cis.poly.edu/~gelley.

**4.2 Features**

We extracted 38 features from each page in our dataset; these are divided into three categories.

The first is metadata about the page. This includes revision features, such as the number of revisions, the number of unique editors who contributed, and whether most of the article's text was written by its creator. This category also includes metrics used to roughly approximate the popularity of the page or notability of its subject [8].These metrics are: the number of Wikipedia pages that link to the article, the number of external web pages that link to it, the number of page views it has received, and the number of Google hits for the page title. (To ensure that the hits were actually relevant to the topic, we enclosed titles of 3 words or less in quotation marks before searching.)

The second category is page creator characteristics. We used Wikipedia's API to retrieve the following information about each page creator: the length of time the user was registered and the number of edits he made before creating the page in question, , and whether the account has been blocked, a disciplinary action generally taken when the user has been guilty of vandalism. We also noted whether the article creator had a userpage, a site-specific personal homepage used extensively by many experienced editors, and how many other pages they had created which were kept. Many of the articles in our dataset are biographical and appear to have been created by the subject of the article; besides for being a clear violation of Wikipedia policy [15], these articles tend to be non-notable. For example, an article about "John D Smith" created by user "JDSmith" is likely autobiographical. We detect some of these articles by using a simple similarity measure to determine the string similarity between the title of the article and the username of its creator.

Finally, we looked at characteristics of the article itself. This includes the number of sections, images, and references it contains, and how many links it contains to other Wikipedia articles. (We normalized the last by article length.) For these, we used the final version of the article, since this is the one that is most complete and the one that human editors look at when evaluating whether an article should be deleted. To get some idea of the quality of the writing, we calculated several reading level metrics for each article, including the Flesch-Kincaid and SMOG indices. We also retrieved the normalized frequencies of nouns, verbs, adjectives, and adverbs in the text. Further details can be found on our website.

**4.3 Analysis**

Using the features described above, we analyzed our dataset to determine whether there were any overall differences between the kept and deleted articles. We first focused on the features related to the article creator, hypothesizing that experienced users are more likely to create notable articles. We found that indeed, there were large differences in creator characteristics between the kept and deleted sets. Creators of kept articles had, on average, been registered on Wikipedia before creating the page for five times longer than those whose pages were deleted, and had made 9 times more edits. In fact, 76% of deleted pages were created the day their creator joined Wikipedia, while only 22% of deleted pages were. Creators of kept pages also had a better track record: 85% of kept pages had a creator who started at least one other page on Wikipedia, and the creators of 61% had created more than 10. In contrast, the corresponding figures for deleted pages were 30% and 4%. Kept pages were also four times more likely to have a creator with a userpage. (See section 4.2) Deleted pages were also three times more likely to have a creator who was later blocked. (See section 4.2). Of the pages that were deleted, 14% were very likely autobiographical (similarity > .5; see section 4.2), while only 1% of the kept pages were.

Other features showed similar disparities. On average, kept pages had nearly 10 times as many revisions as deleted ones. 84% of kept articles belonged to more than two categories, while just 6% of deleted articles did. Only 36% of deleted articles had even one section header, a basic convention that most standard articles have; fully 85% of kept articles did. References, a basic hallmark of notability, showed a similar gap: 34% of kept articles had at least two references, while only 9% of deleted articles did. Google hits for the titles of deleted articles fell mostly into two ranges: wildly enormous, suggesting that the title was composed of common words or phrases and is probably not about a specific encyclopedic topic, or extremely small, probably pages about non-notable people. Google hits for kept pages were mostly mid-range. (We were unable to draw any conclusions from pageviews due to a malfunction in Wikipedia's platform software that undercounted pageviews in December 2011.)

Deleted pages had more adjectives, and twice as many verbs and adverbs as kept ones. We speculate that these parts of speech signify an active, narrative writing style, unlike the more formal tone of an encyclopedia. The kept pages averaged a reading level four grades higher than that of the deleted.

The above comparisons were performed on the original dataset. Given that some of the features can be functions of the age of the article, we repeated the comparisons on our time-normalized datasets. The differences, while smaller, were still significant. The table below contains some representative features and their values for the original and older pages datasets. (See section 4.1)

|                                              | Old Kepts | Old Deleted | All Kepts | All Deleted |
|----------------------------------------------|-----------|-------------|-----------|-------------|
| Total                                        | 1465      | 206         | 1386      | 2496        |
| Creator # edits before creating page (mean)  | 14662     | 1878        | 13980     | 1564        |
| Creator joined site the day the page was created | 20%   | 42%         | 22%       | 76%         |
| Mean days on site before page creation       | 482       | 286         | 531       | 102         |
| Median days on site before page creation     | 216       | 23          | 233       | 0           |
| No references                                | 45%       | 59%         | 44%       | 77%         |
| More than 1 reference                        | 19%       | 14%         | 34%       | 9%          |
| Mean FK reading level                        | N/A       | N/A         | 15.90     | 12          |
| Average # of revisions                       | 55        | 21          | 48.80     | 5           |
| Fewer than 5 revisions                       | 3%        | 10%         | 7%        | 74%         |
| Only has 1 unique editor                     | 0         | 3%          | 4%        | 72%         |
| > 5 unique editors                           | 89%       | 63%         | 83%       | 5%          |
| More than 2 categories                       | 86%       | 50%         | 84%       | 6%          |
| At least 1 image                             | 11%       | 9%          | 10%       | 2%          |
| > 10 WP pages link to here                   | 59%       | 9%          | 55%       | 2%          |
| Likelihood autobio > .5                      | 2%        | 4%          | 2%        | 15%         |
| More than 1 section                          | 85%       | 65%         | 85%       | 36%         |
| Links to other WP pages (normalized; mean)   | 9.2       | 6           | 9.20      | 2.7         |

**Table 1: Comparison of various features for kept and deleted datasets.**

Our analysis also showed that very fast deletion is widespread. 47% of the articles in our Speedy deletion set were nominated for deletion less than 10 minutes after they were created, and 90% were nominated within a day. This was especially pronounced when it came to new users; an article created by a user whose account was opened that day was more than three times more likely to be deleted than one created by a user who had been registered for a year.

## 5. Discussion

From our analysis, it is clear that there are significant differences between pages that are deleted and those that are not, even when the influence of time is accounted for. This difference does suggest that in many cases, the articles deleted are genuinely non-notable. There are, then a large number of article creators who are adding non-notable material to Wikipedia. From the content of the pages, it is clear that they are mostly not vandals, just genuinely unaware of the fact that their subjects did not warrant a Wikipedia article. For a site with extremely detailed notability guidelines, Wikipedia is not doing enough to communicate even the most basic concept of notability to its users.

Existing methods are obviously not working. The Article Wizard, which is intended almost entirely to weed out potential articles that are non-notable, is extremely underutilized. About 300 pages are created every day with the Article Wizard, but over 1,000 are created without it. Wikipedia's how-to essay for creating an article [15] gets as many daily pageviews as there are articles created. From the sheer number of non-notable articles created, however, it is clear that the essay is not effective enough at educating editors about notability.

Some of the problem stems from Wikipedia's own ethos. Wikipedia's goal is to be open-access, easily available for anyone to contribute. The site's eagerness to lower barriers to participation, however, has had the unintended effect of raising the barriers to *effective* participation. By strictly separating policy pages from the main body of the encyclopedia, and requiring no education in site protocol before editing or even page creation, Wikipedia creates multitudes of new editors who are remarkably ignorant of the basic goals and policies of the encyclopedia [4]; lacking enough direction from the site itself as to what sort of articles it wants, they are forced to decide for themselves. Volumes of notability guidelines are useless if the people they are intended for never see them.

Our analysis also uncovered a very large amount of extremely rapid deletions. This haste might be warranted in the case of vandalism, but notability can be very difficult to determine in short, young articles. If given a little more time, it is quite possible that some apparently non-notable articles will provide proof of importance and develop into useful articles. In fact, 7% of the articles in our kept dataset were at some point nominated for Speedy Deletion for lack of notability. All have since developed into standard Wikipedia articles.

## 6. Conclusion and Recommendations

It is clear that the fault for the large numbers of notability-based deletions lies with both the Wikipedians who create the articles and those who delete them. As we have seen, Wikipedia's policies are thoughtful and aim to provide the best possible balance between encouraging the addition of as much noteworthy content as possible and keeping the site free of unencyclopedic articles. These policies, however, are not being implemented in an optimum fashion. Too few novice editors are familiar with the notability guidelines, and too few article deleters take the deletion guidelines seriously enough. If Wikipedia were to make a concerted effort to help all users become familiar with notability policy, it might see the volume of non-notable pages created drop.

As we showed above, the Article Wizard is an underutilized resource. This is understandable, given that all Article Wizard articles must be reviewed before posting, and the review backlog is a week long. The Wizard is also presented as very optional, rather than as a valuable resource. Tinkering with the Article Wizard to allow users who wish to post their articles directly to the article namespace, along with a sitewide campaign to encourage editors, especially new ones, to use it, might help many more users evaluate their prospective pages for notability before creation. Even a small reduction in the number of non-notable articles created would reduce the pressure on the New Page Patrollers; less stress might make them less prone to quick decisions and allow them the time to fully evaluate each page before deciding. If it's combined with a conscious effort to be somewhat less hasty with deletions, fewer promising new editors will be scared away.

One additional possibility to consider is the use of automated tools to assist deleters in making their decisions. Preliminary experiments in using our features to construct a machine learning classifier that can discriminate between kept and deleted pages, even at a young age, are promising. We plan to assign a score to each new page based on its likelihood of developing into a worthwhile article. These scores can assist new page patrollers in deciding whether to delete a page right away or give it some extra time. That extra knowledge can make a big difference in reducing the number of promising articles deleted too hastily, as well as the time and effort spent weeding out those pages that truly do not belong.